\def\h{{$h$${\nu}$}}
\def\op{{$\omega_p$}\/}
\def\os{{$\omega_s$}\/}

\documentclass[prb,preprint,amsmath,amssymb,aps,floatfix]{revtex4-2}
\usepackage{comment}
\usepackage{float}
\usepackage{wasysym}
\usepackage{graphicx}
\usepackage{dcolumn}
\usepackage{bm}
\usepackage{multirow}
\usepackage{array}
\usepackage{booktabs}
\usepackage{amsmath}
\usepackage[normalem]{ulem}
\usepackage{xr}
\usepackage{cleveref}
\usepackage{xcolor}
\usepackage{txfonts}
\newcommand{\srb}{\textcolor{black}}
\pdfoutput=1 

\begin{document}
	\title{Intrinsic and extrinsic plasmons in the hard x-ray photoelectron spectra of nearly free electron metals}

	\author{Mohammad Balal$^1$, Shuvam Sarkar$^1$, Pramod Bhakuni$^1$, \\Andrei Gloskovskii$^2$, Aparna Chakrabarti$^{3,4}$,  Sudipta Roy Barman$^{1*}$ }
	\affiliation{$^1$UGC-DAE Consortium for Scientific Research, Khandwa Road, Indore 452001, India}
	\affiliation{$^2$Deutsches Elektronen-Synchrotron DESY, Notkestrasse 85, D-22607 Hamburg, Germany}
\affiliation{$^3$Raja Ramanna Centre for Advanced Technology, Indore 452013, Madhya Pradesh, India	}  
\affiliation{$^4$Homi Bhabha National Institute, Training School Complex, Anushakti Nagar, Mumbai 400094, Maharashtra, India}
\begin{abstract}	Collective plasmon excitations in solids that result from the process of photoemission are an important area of fundamental research. In this study, we identify a significant number ($n$) of multiple bulk plasmons ($n$\op)  in the hard x-ray photoelectron spectra of the core levels  and  valence bands (VBs)  of two well-known, nearly free electron metals,  aluminum (Al) and magnesium (Mg). On the basis of earlier theoretical works, we estimate the contributions of extrinsic, intrinsic, and interference processes to the intensities of 1$s$ to 2$s$ core level plasmons. The intrinsic contribution  diminishes from  22\%  for 1\op\, to 4.4\% for  2\op, and becomes negligible thereafter (0.5\% for 3\op). \srb{The extrinsic and intrinsic plasmon contributions  
	~do not vary significantly  across a broad range of photoelectron kinetic energies, and also between the two metals (Al and Mg).} The interference contribution varies from negative to  zero as $n$ increases. An asymmetric line shape is observed for the bulk plasmons, which is most pronounced for 1\op.   Signature of the surface plasmon  is detected in normal emission, and it  exhibits a significantly increased intensity in the grazing emission.  The  VB spectra of Al and Mg, which are dominated by $s$-like states, exhibit excellent agreement with the calculated VB based on density functional theory. The VB exhibits four multiple bulk plasmon peaks in the loss region, which are influenced by  an intrinsic  process in addition to the extrinsic process. On a completely oxidized aluminum surface, the relative intensity of the Al metal bulk plasmon remains nearly unaltered, while the surface plasmon is completely attenuated. 
	\end{abstract}
	\maketitle 
\newpage

\section{Introduction}

The investigation of plasmon excitations in photoelectron spectra of free electron metals has been a central focus of fundamental research for over four decades, with ongoing theoretical~\cite{Mandal22,Fujikawa22,Shinotsuka08,Fujikawa08,Zhou18,Kazama14,Yubero05, Hedin03,  Mahan00,Hedin98,Inglesfield83,Inglesfield83a,Bose82,Penn78,Penn77,Baer73,Citrin77,Sunjic74,Langreth74,Mahan73,Chang72,Langreth71}  and experimental~\cite{David16,Ozerprl11, Biswas03,Santana18, Barman04, Barman01, Barman98,Attekum78,Attekum79, Baird78, Steiner78,Hoechst78,Bradshaw77,Fuggle76, Pardee75} investigations. 
~These investigations have provided valuable insights into the physics of plasmons   in the nearly free electron metals. Furthermore,  continuous interest in this field highlights its relevance and potential for further advancements in our understanding  of collective electronic excitations and  the development of various future applications, such as plasmonics. The bulk and surface plasmons -- observed as loss features in the core level photoelectron spectra -- are dominated by an extrinsic process that results from the Coulomb interaction of the  conduction electrons with the photoelectron traversing through  the solid from the photoemission site to the surface.  An additional contribution to the plasmon intensity comes from an intrinsic process, where  the sudden change in the potential  due to the formation of a localized core hole attracts the  conduction electrons  to screen it, resulting in a plasmon oscillation.   From the variation of successive multiple bulk plasmon ($n$\op)  intensity in Al 2$s$ x-ray photoemission spectrscopy (XPS), the intrinsic plasmon was estimated to be  14-25\% of the total  intensity~\cite{Attekum78,Steiner78}.  Furthermore, an interference~\cite{Inglesfield83,Inglesfield83a,Langreth71,Attekum78,Bose82,Biswas03} between the intrinsic and extrinsic processes, where virtual plasmons created by one are  absorbed by the other,  plays an important role in determining the shape of the plasmons.   Hedin~\cite{Hedin03} pointed out that  the interference term should decrease as a function of the photoelectron kinetic energy. The coupling of  the Al surface plasmon to the oxygen adsorbed on its surface was observed in the O 1$s$  core level spectra in XPS~\cite{Bradshaw77}.   Biswas \textit{et al.}~\cite{Biswas03} determined the relative contributions of the intrinsic, extrinsic, and interference processes in the  surface plasmon  from the XPS core level spectra of Al metal. The loss region has been studied for  Al and carbon  by XPS to identify both the one electron interband transitions and plasmon excitations ~\cite{David16,Santana18}. Bulk plasmons have  been observed in the XPS valence band (VB) spectra  of  Al~\cite{Attekum78} and Mg up to $n$= 2-3~\cite{Attekum79,Hoechst78},  \srb{but their intensities were affected by presence of KLV Auger electron spectroscopy signal and x-ray satellites.} The mechanism of instrinsic plasmon production, where, unlike the core level, the photo-hole is delocalized, was related to many-body effects and electron-electron interaction by theory~\cite{Penn78,Bose82}. 
 
Hard x-ray photoelectron spectroscopy (HAXPES) has become a vital method to study the bulk electronic structure of materials because of its large inelastic mean free path (8-10~nm)~\cite{Kalha21,Kobayashi09,Fadley2010,Woicik16, Gray,Gray_a,Nayak12,Ohtsuki11,Takata08,Sing09, Sarkar21,Sarkar21a,Sadhukhan19,Sadhukhan23}.  Previous research has investigated this phenomenon for a limited number of elements, including carbon~\cite{Kunz09}, silicon~\cite{Offi07}, germanium~\cite{Novak07},  phosphorous~\cite{David20}, and  Al films grown on Si~\cite{Konishi23}.  However, the investigation of plasmon excitation in nearly free electron metals remains a captivating subject due to the substantial theoretical research that has been undertaken in this area previously~\cite{Mandal22,Zhou18,Kazama14,Shinotsuka08,Fujikawa08,Yubero05, Hedin03,  Mahan00,Hedin98,Inglesfield83,Inglesfield83a,Bose82,Penn78,Penn77,Baer73,Citrin77,Sunjic74,Langreth74,Mahan73,Chang72,Chang72a,Langreth71}.
Notwithstanding this, theoretical investigations of  plasmons in free electron metals in the HAXPES regime are scarce. 
Shinotsuka \textit{et al.}~\cite{Shinotsuka08} performed a full quantum mechanical calculation, taking the interference term into account, using the model suggested by Hedin \textit{et al.}~\cite{Hedin98}. Fujikawa \textit{et al.} have attempted to address this issue through the application of the quantum Landau formula, which accounts for elastic scatterings both prior to and subsequent to the loss~\cite{Fujikawa08}. These works show that in the HAXPES regime, the extrinsic contribution  is dominant, although  both  the positive intrinsic   and the negative interference  terms are present. The interference and intrinsic contributions decrease very slowly as a function of the photoelectron energy and  are therefore  significant even at a large photon energy of 10 keV. Even the surface plasmon exhibits a peak of significant intensity at this photon energy~\cite{Fujikawa08}.  The other interesting theoretical result~\cite{Shinotsuka08} is that  the loss features  hardly exhibit any angular dependence i.e., the grazing versus the normal emission spectra are nearly similar. This behavior is different compared to the low photon energy XPS regime~\cite{Biswas03,Yubero05,Inglesfield83,Inglesfield83a,Shinotsuka08}. The above-discussed intriguing characteristics of the plasmons, as postulated by theory~\cite{Fujikawa08,Shinotsuka08}, have yet to be empirically investigated. 

\srb{HAXPES study of the core-level and VB plasmons in  free electron metals is scarce in previous literature, } despite the fact that this technique  offers unique advantages over XPS due to its deeper probing depths.  \srb{ Another benefit of HAXPES over XPS is that the former does not  experience interference from the  Auger electron spectroscopy signal.} 
\srb{A recent study on Al thin films on Si  and that of fractured Al metal reported  Al 1$s$ related plasmons~\cite{Konishi23}.  However, the Si(111) surface was naturally oxidized and  the ex-situ introduced Al films were inhomogeneous  and showed a  large Al-oxide component  in the Al 1$s$ spectrum. Thus, although bulk plasmons for clean bulk Al were observed,  the thin film related Al plasmons were affected by presence of oxide and Si related loss features. }  

Here, we present a detailed HAXPES study of  the plasmon excitations in the VB and core level spectra for clean Al and Mg bulk single crystals.  A large number of multiple bulk plasmons ($n$\op) up to $n$= \srb{13} is observed, in addition to a relatively weaker surface plasmon (\os). We conduct exhaustive multi-parameter curve fitting considering  the plasmons to ascertain the variation of their  intensity and shape. From the intensity variation,  the intrinsic, extrinsic, and interference contributions to $n$\op\, are determined based on prior theoretical works~\cite{Langreth74,Langreth71,Chang72,Chang72a,Penn78}.  Normal and grazing emission properties of the bulk and surface plasmons have been investigated and compared with an oxidized Al surface.  In addition, our work shows  the existence of the  intrinsic process in the VB \srb{related bulk} plasmons.

\section{Methods}
The experiments were carried out at the P22 beamline in PETRA III, Deutsches Elektronen-Synchrotron, Hamburg, Germany. A post-monochromator was used to improve the resolution and stability of the photon beam. All the spectra were measured using 6 keV photon energy. The electron energy analyzer has an angular acceptance angle of $\pm$15$^{\circ}$.  The overall energy resolution, including the source and the  analyzer contribution, was set at 0.3-0.4 eV, as measured from the Au Fermi edge  that was in electrical contact with the specimens. The details of  the beamline and the endstation can be found in Ref.~\onlinecite{P22}.  The measurements were carried out at $\theta$=  85$^{\circ}$ (referred henceforth as normal emission) and 10$^{\circ}$ (grazing emission), where $\theta$ is the angle between the  analyzer axis and sample surface.  
  
  Polished and oriented Al single crystals were cleaned \textit{in-situ} by sputtering with  Ar$^{+}$ ions of energy 3-5 keV for the core-level studies.   In the case of the VB  and some core-level studies, the Al and Mg crystals were mechanically scraped \textit{in-situ} with a diamond file.   
    ~The binding energy ($E$)  scale of both the core level and valence band spectra has been  defined from the Fermi level position of Al/Mg metal (rather than from Au) because of the recoil shift~\cite{Takata08,Fadley2010}; this is discussed later in subsection IIIC. The HAXPES core level main peaks have been fitted using  the least squares error method, where the Doniach-\v{S}unji\'c (DS) line shape represents the main peak~\cite{Doniach70}.  
   ~Asymmetric Lorentzian line shapes have been used to represent the plasmon loss peaks~\cite{Biswas03,Attekum78}, as also suggested by the theoretical 1\op\, and 1\os\, line shapes from Ref.~\onlinecite{Shinotsuka08} shown in Fig.~S1 of Supplementary Material (SM)~\cite{supple}. The lifetime broadening of the core level main peak, the DS asymmetry parameter, intensities, peak positions, and the inelastic background have been varied. \srb{The half width at half maximum of the life-time broadening  and the DS asymmetry parameter for Al (Mg) 2$s$, determined from our fitting, are 0.4 (0.32) eV and  0.1 (0.12), respectively.} Each of the  plasmon peaks has been simulated by four parameters: intensity, position,  and an asymmetric Lorentzian~\cite{Lorentzian} with independently varying left ($\Gamma_L$) and right ($\Gamma_R$) half width at half maximum.  \srb{Besides the $n$\op\ peaks, 2\os, 1\op+1\os, and 2\op+1\os\ surface plasmons  were also considered for Al 1$s$.}   \srb{The inelastic background was simulated by the Tougaard method~\cite{Tougaard89,Tougaard89a}, whereas use of the Shirley method resulted in  unphysically large value of the intrinsic plasmon probability}. 
 ~As in our earlier works~\cite{Biswas03,Sadhukhan19a}, all the parameters were not varied simultaneously to avoid unphysical solutions: for example,  the parameters for  plasmons with  $n$$\geq$7 were varied individually. Thus, the  number of  independently variable parameters for  1$s$  (2$s$) was  \srb{68 (39). } 

The calculation of the electronic properties of the bulk Al and Mg metals has been performed using the all-electron full potential linear augmented planewave (FPLAPW) method, including scalar relativistic corrections, by employing the density functional theory (DFT) based WIEN2k\cite{wien} program package. For the exchange-correlation functional, we consider the generalized gradient approximation over the local density approximation, given by Perdew, Burke, and Ernzerhof (PBE)~\cite{pbe}. Energy cut-off values of about 15 Ry have been used in the calculations. The charge density cut-off is 14. Further, we have used 10 as the maximum value of angular momentum for the ($l, m$) expansion of wave function and density. The convergence criteria for energy and charge have been taken to be 10$^{-5}$ Ry and 0.001 e$^{-1}$, respectively. The total density of states (DOS) is calculated from the Kohn-Sham orbitals, and its integration up to the Fermi level gives the total number of valence electrons. The partial DOS (PDOS) is obtained by projecting the total DOS for each orbital ($s$, $p$, $d$). For Al and Mg with delocalized states,  touching muffin-tin sphere was taken to minimize the discrepancy between the total DOS and the sum of all the PDOS.

To calculate the theoretical VB, the angular momentum  projected  PDOS was multiplied by the corresponding photoemission cross sections per electron calculated for 6 keV  photon energy by Trzhaskovskaya \textit{et al.}~\cite{Trzhaskovskaya18}.  In the case where an energy level is not occupied and so the cross section is not calculated, the closest value is assumed, as in our previous work~\cite{Sadhukhan19,Sadhukhanthesis}. For example, for Mg $p$ PDOS, the cross-section is taken to be the same as that of Al $p$.   Furthermore, the partial contributions are  added and multiplied by the Fermi function  and subsequently convoluted with a Gaussian function representing the instrumental resolution. Additionally, an energy dependent Lorentzian broadening given by  $\mu\times$$E$ has been used, where $E$ is the binding energy and $\mu$ is taken to be  0.15 eV~\cite{Fujimori84,Barman95}. 

\begin{figure*}[tb]
	\centering
	\includegraphics[width=0.85\textwidth,keepaspectratio,trim={2.7cm 1cm 1.5cm 0cm },clip]{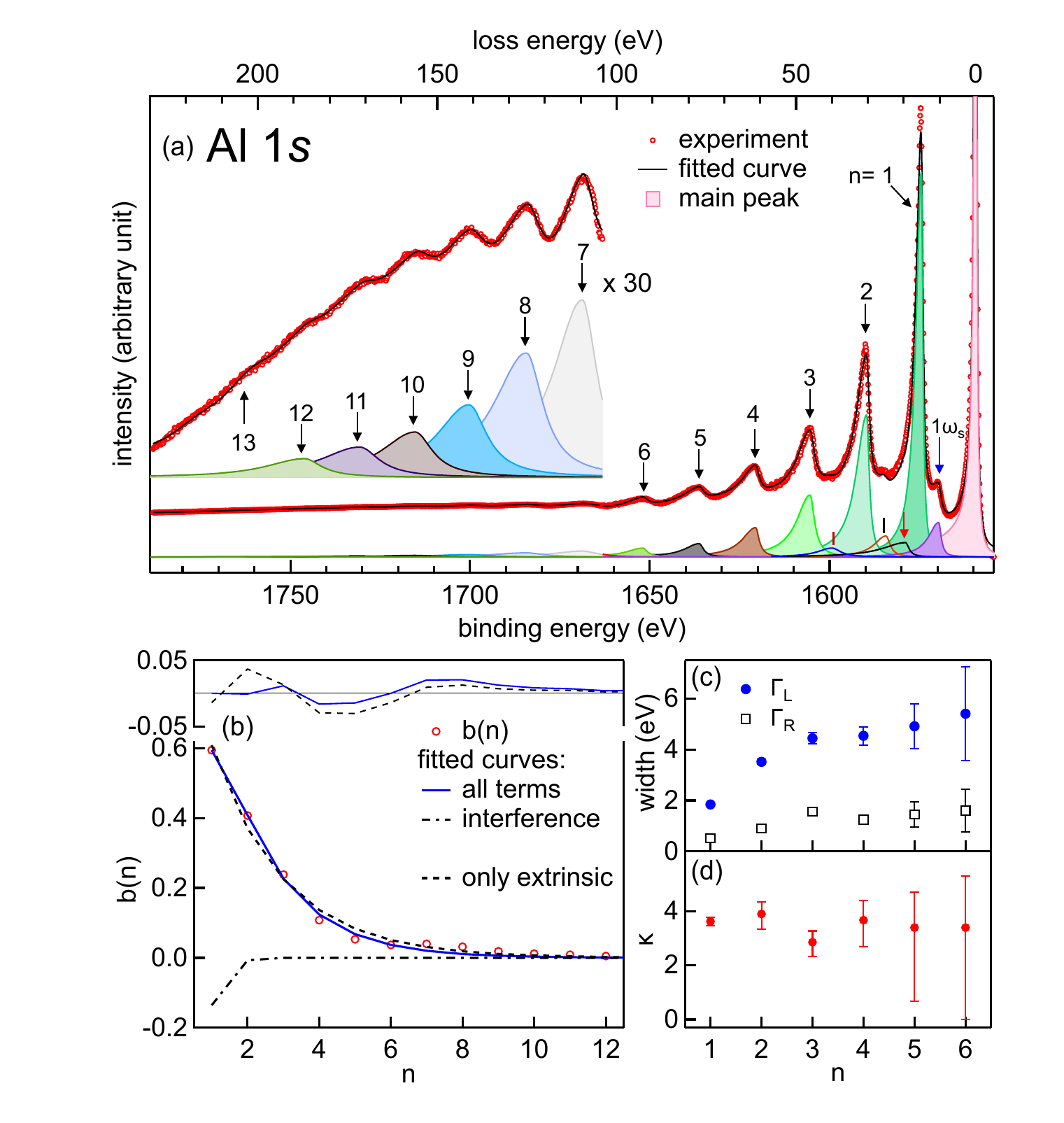}
	\caption{(a) Al 1$s$  core level spectrum (red open circles) of Al  taken at  normal emission ($\theta$= 85$^{\circ}$) showing multiple bulk plasmon peaks $n$\op~($n$= 1-\srb{13}, black arrows). 
		~The loss energy scale (the top horizontal axis) is defined with respect to the main peak. The latter is truncated to show the plasmon peaks on an expanded scale.  The spectrum  for $n$= 7 to \srb{13}  is shown separately in a magnified scale. The fitted curve (black) and each of the $n$\op~ peaks are shown. 	 (b) The normalized intensity of the bulk plasmons ($b(n)$)  fitted with Equation 1 that includes all the terms i.e., extrinsic, intrinsic, and interference (blue curve) and with only the extrinsic term (black dashed curve). The corresponding  residuals of the fitting are shown in the top panel in same line type. (c) $\Gamma_R$, $\Gamma_L$, and (d) $\kappa$ (= $\Gamma_L$/$\Gamma_R$) as a function of $n$. }
	\label{bulkplasmon}
\end{figure*}

 \section{Results and Discussion}
 \subsection{Multiple bulk plasmons in Aluminum core level spectra}
\begin{figure*}[tb]
	\centering
	\includegraphics[width=\textwidth,keepaspectratio,trim={1cm 0.5cm 1cm 0cm },clip]{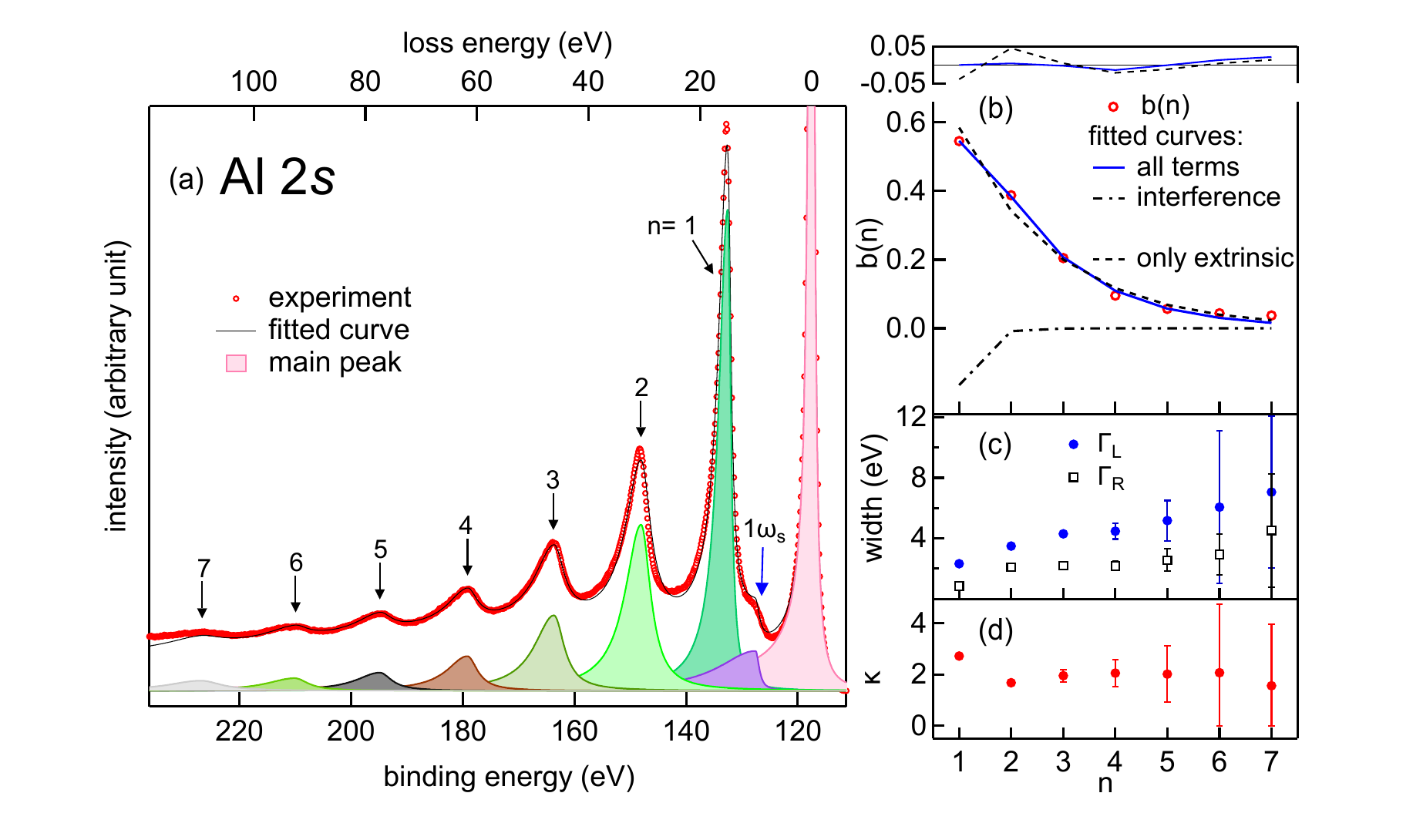}
	\caption{(a) Al 2$s$ core level spectrum taken  at  normal emission showing multiple bulk plasmon peaks $n$\op~ ($n$= 1-7, black arrows) and a surface plasmon peak (1\os, blue arrow).   The  black curve through the data points (red open circles) 	and  each of the $n$\op~ peaks (shaded by different colors) are obtained from the least squares curve fitting.  (b) \srb{$b(n)$ fitted with Equation 1  (blue curve) and with only the extrinsic term (black dashed curve), the respective residuals of the fitting are shown in the top panel.} (c) $\Gamma_L$, $\Gamma_R$, and (d) $\kappa$ (=$\Gamma_L$/$\Gamma_R$)  as  function of $n$.}
	\label{bulkplasmonAl2s}
\end{figure*}

\noindent{\underline {Al 1$s$}:} The Al 1$s$ core level spectrum  in Fig.~\ref{bulkplasmon}(a)  shows multiple bulk  plasmon  peaks ($n$\op) with $n$ as large as 13, as is evident from the magnified spectrum. These plasmons appear on the higher binding energy ($E$) side i.e. the loss region of the main peak at  1559.3~eV.   The first bulk plasmon   peak (1\op)  appears at  \op\  loss energy, and the subsequent $n$\op\ multiple  plasmons appear at  equal \op\, intervals. The positions of the plasmon peaks thus provide an average estimate of \op\ to be 15.3~eV; a similar value is obtained from Al 2$s$ plasmons in Fig.~\ref{bulkplasmonAl2s}(a). 
The reasons that large number of multiple bulk plasmons could be observed are  large photoemission cross-section of  the $s$ states  at high photon energy (6 keV),   the large flux of the photon beam,  the $n$= 7-13 plasmon region recorded with larger dwell time, and   the absence of any other   photoelectron or Auger peaks 
~that could obfuscate the plasmons.  The 1\os\ surface plasmon  is also distinctly visible, as shown by the blue arrow.  The behavior of this surface plasmon will be discussed further in subsection III\,B.    
\srb{Bulk plasmons up to $n$= 7 and the 1\os\ surface plasmon the 1$s$ spectrum of Al metal were reported recently~\cite{Konishi23}, but quantitative analysis their shapes and intensities were not performed. }

Here, based on rigorous multi parameter curve fitting,   the normalized intensities [$b(n)$, i.e., area under $n$\op\ divided by that of the main peak] and shapes [left ($\Gamma_L$) and right ($\Gamma_R$) half width at half maximum of an asymmetric Lorentzian~\cite{Lorentzian}] have been determined. \srb{Besides taking a realistic shape of the bulk plasmons justified by theory in subsection III\,B, inclusion of  weak surface plasmon contributions such as  2\os (red arrow), 1\op+1\os\ (black tick), and 2\op+1\os\ (red tick), improves the quality of the fitting. } We find that $b(n)$ determined \srb{from the area of the plasmons  obtained from fitting (shaded with different colors)}  decreases with $n$, as  shown by the red open circles in Fig.~\ref{bulkplasmon}(b). This variation can be explained by Equation~1, as discussed later. Figure~\ref{bulkplasmon}(c)  portrays the asymmetric shape of  $n$\op,  where  $\Gamma_R$  is smaller compared to $\Gamma_L$. Both show a nearly similar  increasing trend  with  $n$.  The  larger width  of $(n+1)$\op\,   compared to $n$\op\, is because the former is excited by the latter and  can be regarded as the self-convolution of the latter~\cite{Pardee75}.  The asymmetry can be  quantified by the ratio $\Gamma_L$/$\Gamma_R$ (= $\kappa$), its variation with $n$ is shown in  Fig.~\ref{bulkplasmon}(d). 

\noindent{\underline {Al 2$s$}:} Figure~\ref{bulkplasmonAl2s}(a)  shows the Al 2$s$ spectrum   at  normal emission, 
~showing multiple plasmon peaks $n$\op\ up to $n$= 7 \srb{(compared to $n$= 13 for Al 1$s$) because of the relatively lesser photoemission cross section of  2$s$  compared to  1$s$. The intensity of the former is $\sim$9 times less, as shown by the Al survey spectrum in Fig.~S2 of SM~\cite{supple}. For the same reason, although 1\os\  surface plasmon is also distinctly visible at 10.4 eV loss energy (blue arrow),  the multiple surface plasmons are not apparent.}  Another reason for smaller $n$ is that  Ar 2$p$ and 2$s$ peaks  at $E$= 242.2 and 320.3 eV, respectively  (Fig.~S2 of SM~\cite{supple})  interfere because  a small amount of  Ar is implanted  during  the sputtering  process~\cite{Biswas04,Dhaka08}. 
~In particular, Ar 2$p$ peak  coincides with the 8$^{\rm th}$ plasmon, and the former has  a larger intensity. Additionally, the Al related plasmons excited by the Ar 2$p$ photoelectrons~\cite{Dhaka10}  overlap with   $n>$7 plasmons.   
 $b(n)$ determined from multi parameter curve fitting is shown in Figure~\ref{bulkplasmonAl2s}(b). Figure~\ref{bulkplasmonAl2s}(c) shows that the asymmetry of the Al 2$s$ plasmons  is qualitatively similar to that of the Al 1$s$~:  $\Gamma_R$ is smaller than  $\Gamma_L$, and both increase with $n$.  The asymmetry in the plasmon line shape given by  $\kappa$  decreases  from about 2.7  ($n$= 1)  to about 1.8 for  $n$$\geq$2-3. 

\noindent{\underline {Bulk plasmon intensity variation}:} The decrease of $b(n)$  with $n$ has been studied theoretically by different  groups~\cite{Chang72,Chang72a,Langreth71,Langreth74}  to ascertain the  contributions of the different processes connected to the bulk plasmon. Based on perturbation theory arguments,  considering only the extrinsic process, it was suggested that  $b(n)$ should vary as $\alpha^n$, where $\alpha$ is a measure of the extrinsic plasmon probability~\cite{Chang72,Chang72a,Langreth71,Langreth74}.  But the black dashed curve that is obtained using this expression shows a large systematic deviation, \srb{as shown by the residual  in the top panel of Fig.~\ref{bulkplasmon}(b) in same line type.} This shows that, in addition to the extrinsic process, the  intrinsic and interference processes could also be significant. So, a modified Langreth equation~\cite{Biswas03,Langreth71,Chang72,Chang72a} that considers all three terms  has been used to represent $b(n)$ as follows:
\begin{equation*} 
	b(n) = \alpha^n\sum_{m=0}^{n} \frac{[(\beta-c\chi)/\alpha]^m}{m!}~~~~~~~~~~~~~~~~~~~~~~~~~~~~~~~~~~~~~~~~~~~~~~~~~~~~~~~~~~~~~~~~~(1)
	\label{modlangreth}
\end{equation*}
where  $\beta$ is the measure of the probability of  the intrinsic process for the $n^{\rm th}$ plasmon~\cite{beta}.  $\chi$ is the probability of the interference process~\cite{Chang72,Chang72a} given by the product of intrinsic and extrinsic plasmon probabilities~\cite{chi}, where  $c$ is the proportionality constant~\cite{Biswas03}. If the above expression is used to fit the experimental $b(n)$ by varying $\alpha$, $\beta$, and $c$,  the quality of the fitting (blue curve)  improves substantially, which is evident from a  reduction in \srb{chi-square error} 
by a factor of  about \srb{2.5} and  comparison of  the  residuals (blue solid and black dashed lines)  in the  top panel.  From the fitting, we obtain  $\alpha$=  \srb{0.54$\pm$0.05} 
~and $\beta$=  \srb{0.19$\pm$0.1.} 
~The experimental value of $\alpha$ is in good agreement with the  theoretically suggested value  of 0.5 by Chang and Langreth~\cite{Chang72,Chang72a}.  The interference term (black dot-dashed curve) is found to be finite but negative, with a rapidly decreasing contribution that becomes nearly zero for $n$$>$3.  

 $b(n)$ for Al 2$s$ has been fitted using Equation~1, as shown by the blue curve in Fig.~\ref{bulkplasmonAl2s}(b). The residual in the upper panel demonstrates the good quality of the fitting. On the other hand, the fit is worse if  only the extrinsic term is considered. From the fitting, 
~we find that   $\alpha$=  \srb{0.53$\pm$0.05}
~and $\beta$= \srb{0.18$\pm$0.1}. 
~Thus, the extrinsic and intrinsic  contributions are similar between  2$s$ and 1$s$,  although for the former,  the kinetic energy ($E_k$) of the electrons is larger (5831 eV) compared to the latter (4390 eV).  
Al 2$s$ recorded by XPS, where $E_k$ is relatively smaller (1133 eV), gave $\alpha$ (= 0.46 eV) and $\beta$ (= 0.22 eV)~\cite{Biswas03} that are not significantly different from the HAXPES values.  This shows that, contrary to the expectation that the extrinsic process might be dominant at larger $E_k$ due to larger inelastic mean free path of the photoelectrons, all  three processes have nearly similar contributions over a large range of $E_k$ from  1133 eV ($\approx$ 1 keV) to 5831 eV ($\approx$ 6 keV). This is  in agreement with the theoretical study by Shinotsuka \textit{et al.}~\cite{Shinotsuka08} that shows hardly any change in the different processes in the 1\op\, bulk plasmon  of Al 
~in the \h\, range of 2 to 5 keV. 

The interference term (black dot-dashed curve) also shows similar trend. Its contribution at $n$= 1 is also nearly similar:  \srb{-0.14$\pm$0.05} 
~for  1$s$ and \srb{-0.17$\pm$0.05}  for  2$s$ using HAXPES, while it is  reported to be -0.2 for 2$s$ from XPS~\cite{Biswas03}. This is in agreement with the calculation by Shinotsuka \textit{et al.}~\cite{Shinotsuka08}, who found that the interference term does not decrease in the HAXPES regime, in disagreement with the prediction by Hedin~\cite{Hedin03}. 

\begin{figure*}[tb]
	\centering
	\includegraphics[width=\textwidth,keepaspectratio,trim={2cm 0.4cm 1cm 0.5cm },clip]{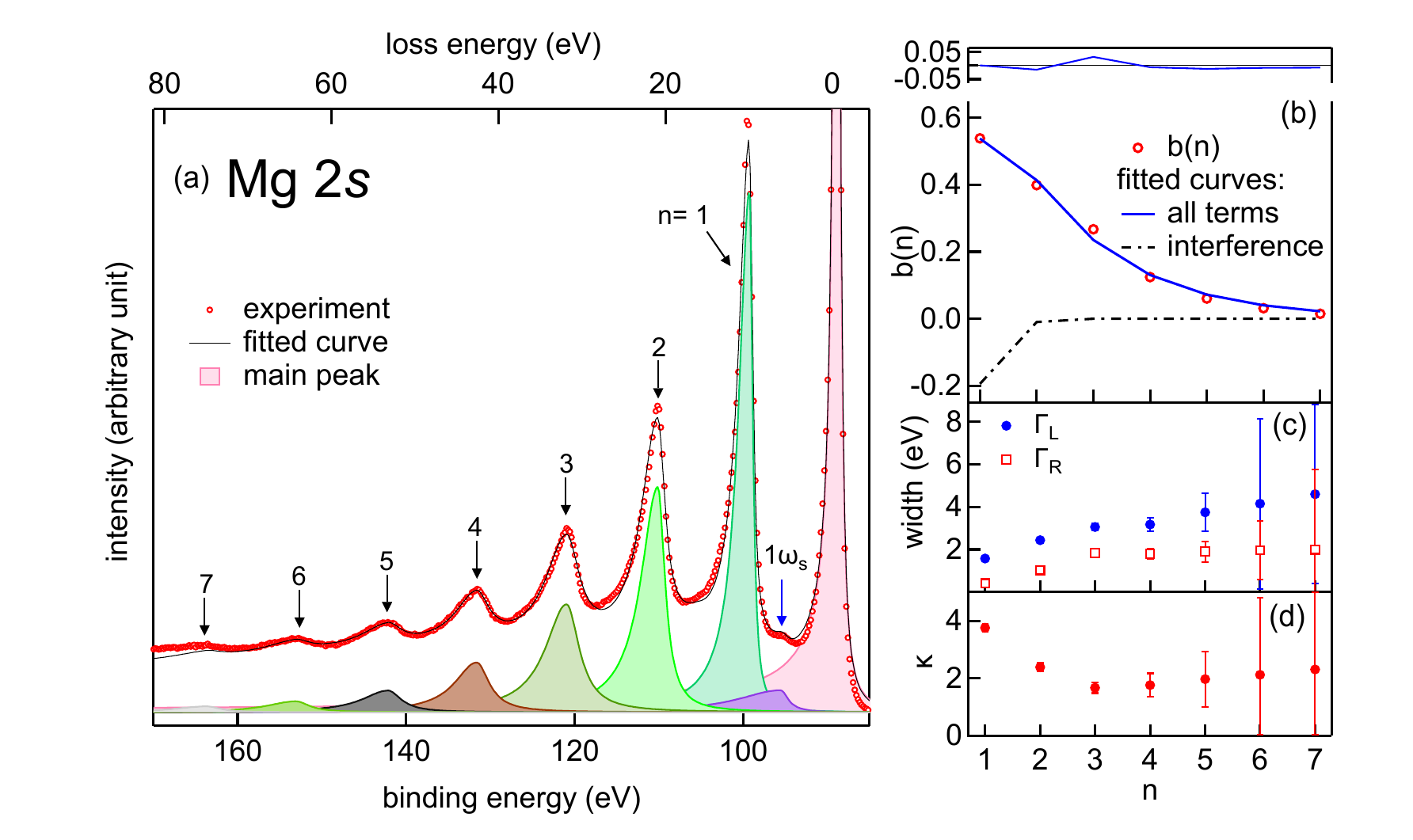}
	\caption{(a) Mg 2$s$ core level spectrum taken  at  normal emission  showing multiple bulk plasmon peaks $n$\op~ ($n$= 1-7, black arrows) and a surface plasmon peak (1\os, blue arrow, also shown in the inset).  The  black curve through the data points (red open circles) and  each of the $n$\op~ peaks (shaded by different colors) are obtained from the least squares curve fitting.  (b) $b(n)$  fitted with all three terms [extrinsic, intrinsic, and interference, (blue curve)]. The residual of the fitting is shown in the top  panel. (c) $\Gamma_L$, $\Gamma_R$, and (d) $\kappa$ (=$\Gamma_L$/$\Gamma_R$)  as   function of $n$.}
	\label{Mg2s}
\end{figure*}

\subsection{Multiple bulk plasmons in  Magnesium core level spectra}
Mg 2$s$ spectrum of clean Mg metal at normal emission, as depicted in Fig.~\ref{Mg2s}(a), also exhibits multiple plasmon peaks denoted by $n$\op\ ($n$= 1-7). \op\ of Mg, determined from the average separation of the plasmon peaks from the main peak, is 10.6~eV.  It is considerably smaller  because of the smaller electron number density of Mg, which  has  2 electrons in the outer shell  compared to  3 electrons in Al.  A weak surface plasmon  is  observed  at 7.5 eV loss energy  (blue arrow). 
~Using curve fitting as in case of Al,  $b(n)$ for Mg has been determined [Fig.~\ref{Mg2s}(b)]. The residual (blue curve) in the top panel of  Fig.~\ref{Mg2s}(b) shows  good quality of the fitting using Equation~1. 
The values of   $\alpha$  and $\beta$   turn out to be 0.55$\pm$\srb{0.05}
~and \srb{0.18$\pm$0.1},  
~respectively.   \srb{ These values are nearly similar to Al. }
~The interference term for Mg (black dot-dashed curve), as in the case of Al, is negative for $n$= 1 and  decreases to almost zero for $n$$\geq$ 3. Figure~\ref{Mg2s}(c) shows that the asymmetry of the Mg 2$s$ plasmons  given by $\kappa$  decreases  from about 4  ($n$= 1)  to below 2 for  $n$= 3~[Fig.~\ref{Mg2s}(d)]. This was also observed for Al 2$s$.  Larger $\kappa$ i.e., asymmetry in the 1\op\, plasmon shape  compared to  $n$$\geq$2 is because of the sizable  intrinsic and interference plasmon contributions  that are more asymmetric in shape compared to the extrinsic plasmon~\cite{Yubero05}.

The $\alpha$ and $\beta$ values obtained for Mg 2$s$ are nearly similar to those from Al 1$s$ and 2$s$, average representative values for these free electron metals being $\alpha$= 0.54$\pm$0.05 and  $\beta$=  0.18$\pm$0.1.
Considering these values, the ratio of  the  intrinsic to extrinsic plasmon probabilities ($e^{-\beta}\beta$:$\alpha$ from Equation~1) for 1\op\,  turns out to be 0.28:1. 
~For 
~2\op\,   the probability ratio \srb{($e^{-\beta}\beta^2/2!:\alpha^2$)} is  0.05:1, 
~whereas for 3\op\  it is  0.005:1 \srb{($e^{-\beta}\beta^3/3!:\alpha^3$)} .  
~Thus, the  intrinsic plasmon contribution decreases rapidly  with $n$ from  22\%  ($n$= 1) to 4.4\% ($n$= 2), and becomes negligible thereafter (0.5\% for $n$= 3). 

\begin{figure}[t!]
	\centering
	\includegraphics[width=0.7\textwidth,keepaspectratio,trim={2cm 0.3cm 1.5cm 0cm },clip]{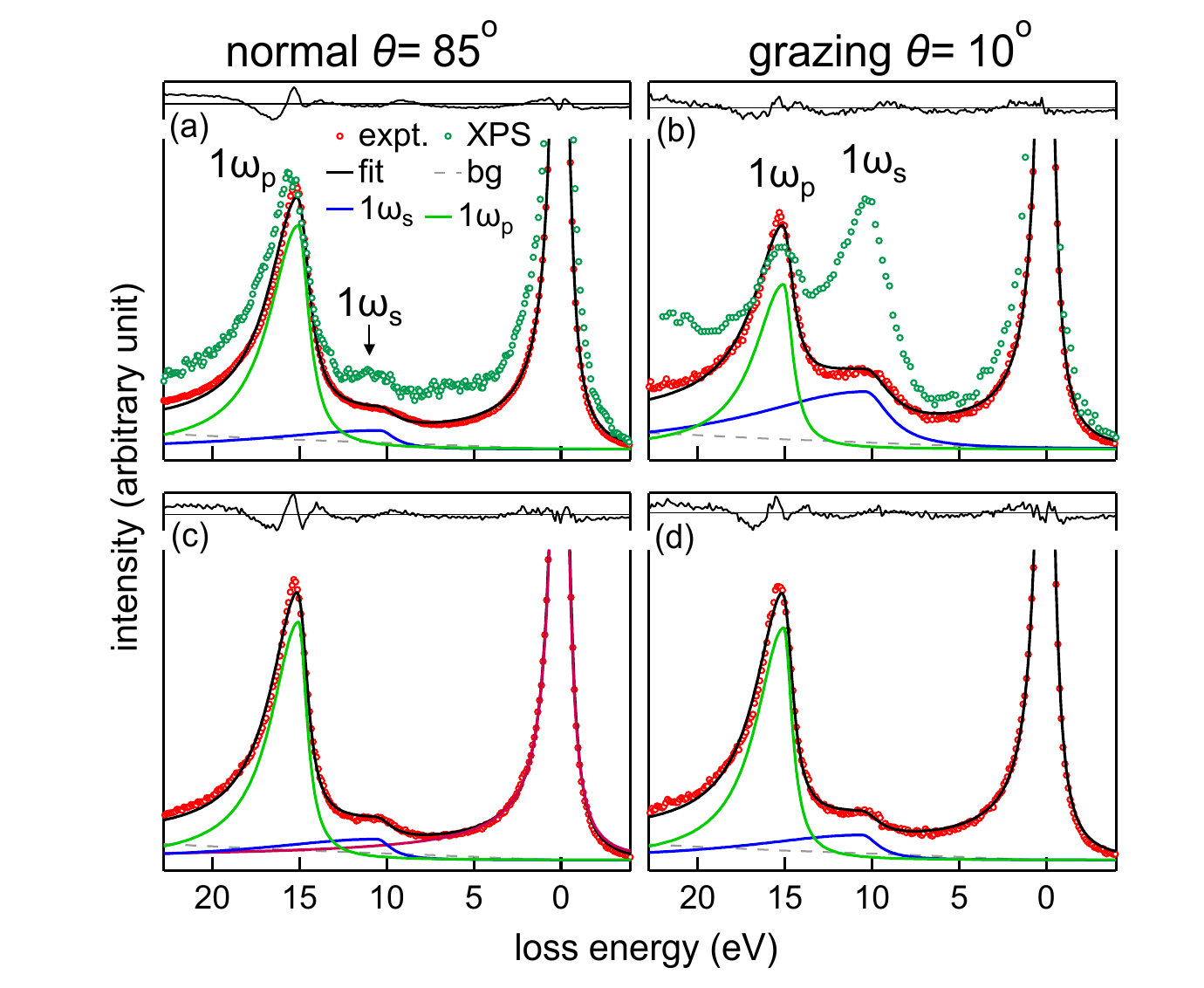}
	\caption{Normal  ($\theta$= 85$^{\circ}$) and   grazing emission ($\theta$= 10$^{\circ}$)  (a, b) Al 2$s$ HAXPES core level spectra showing the  surface plasmon peak (1\os) peak along with 1\op\   for  polished smooth surface compared with XPS data from Ref.~\onlinecite{Biswas03}.   (c, d) Al 2$s$ HAXPES spectra for scraped rough surface in normal and grazing emission, respectively.  The fitted curves including the plasmon shapes,  the inelastic background (bg) and the residual of fitting (top panel) are shown. }
\label{surfaceplasmon}
\end{figure}

\subsection{Surface plasmon behavior  in the Al 2$s$ core level spectra}
\srb{This subsection is dedicated to an in-depth analysis of the surface plasmon.   Figure~\ref{surfaceplasmon}(a, b) shows the 1\os\ along with 1\op\ in Al 2$s$ spectra of  a polished surface cleaned by Ar ion sputtering}. 
~Comparison of the 1\os\, intensity  between normal emission in panel \textbf{a} and grazing emission in panel \textbf{b} shows  a sizable enhancement in the latter.   From the curve fitting, we quantify this by the normalized intensity of 1\os\,  i.e., $s(1)$ (the intensity of  1\os\  divided by the main peak). $s(1)$ increases from 
~\srb{0.13} 
~in normal to \srb{0.44} 
~in grazing emission. \srb{However, a direct comparison with the XPS spectra from Ref.~\onlinecite{Biswas03} [Figs.~\ref{surfaceplasmon}(a,b)] shows the surface plasmon is weaker in HAXPES for both normal and grazing emission. This is  ~attributed to 
	diminished extrinsic contribution  due to larger velocity ($v$) of the photoelectron~\cite{Biswas03, Bradshaw77}.} 
From Fig.~\ref{surfaceplasmon}(a, b),  $b(1)$ decreases from \srb{0.6} 
~in normal  to \srb{0.43} 
in grazing emission, which was related to the variation of extrinsic bulk plasmon contribution with emission angle in  theory~\cite{Baird78}.  The intensity ratio $R$  (= 1\os/1\op) is larger in the grazing \srb{(1 i.e., the surface plasmon is  as intense as the bulk plasmon)} 
compared to normal emission  \srb{(0.2)}. 



The considerable increase of  $s(1)$ 
and $R$  
~in the grazing emission  discussed above is in disagreement with the theoretical results of  Shinotsuka \textit{et al.}~\cite{Shinotsuka08} for HAXPES, where hardly any difference in the calculated  1\op\, and 1\os\, 
~was observed between normal and grazing emission (Fig.~S1 of SM~\cite{supple}).  Nevertheless, to obtain an estimate of $R$ from theory, we have fitted the theoretical spectrum  using asymmetric Lorentzian shapes for  both 1\op\, and 1\os\ from   Ref.~\onlinecite{Shinotsuka08}. A reasonably good  fit of the theoretical shape~\cite{Shinotsuka08} with the asymmetric Lorentzians in Fig.~S1 of SM~\cite{supple} \textit{post-priori} justifies  the use of the latter  to fit the experimental plasmon shapes in Figs.~\ref{bulkplasmon}-\ref{surfaceplasmon}.  From the area under the fitted curves, the theoretical value of $R$ turns out to be close to 0.2, unchanged  for both normal and grazing emission. While $R$= 0.2 is in excellent agreement with the experimental normal emission value \srb{(0.2)}, it is   conspicuously underestimated  for the grazing emission,  where the experimental value  is \srb{1}.
 Enhancement of $s(1)$ in grazing emission has also been reported for XPS. For example,  in Al 2$s$  from XPS, a ten times increase in $s(1)$ has been reported in the grazing emission, which  was explained by an increase in $v$ parallel to the surface~\cite{Biswas03,Bradshaw77}. This effect was possibly not considered in Ref.~\onlinecite{Shinotsuka08}. Note that in Fig.~\ref{surfaceplasmon}, the shape of the surface plasmon is found to be highly asymmetric, with $\Gamma_R$ considerably smaller than $\Gamma_L$. Their ratio $\kappa$ is \srb{4.9 (7.9)} in grazing (normal) emission (Fig.~\ref{surfaceplasmon}). 
~The asymmetric shape of the surface plasmon as well as the bulk plasmons is in agreement with  Ref.~\onlinecite{Shinotsuka08} and the earlier theoretical works~\cite{Inglesfield83,Inglesfield83a,Bose82}. 

Figures~\ref{surfaceplasmon}(c,d) show an interesting result that for rough surface obtained by scraping the polished surface with a diamond file,   the enhancement of the surface plasmon in the grazing emission is not as pronounced: $s$(1)= \srb{0.14} in the normal and \srb{0.17} in grazing emission.  \srb{This is because  the emission angle is not well defined for the rough surface.}  
\srb{Comparison of Figs.~\ref{surfaceplasmon}(a,c) shows that both the bulk and surface plasmon intensities in normal emission are however nearly similar between the smooth [$b(1)$= 0.6, $s(1)$= 0.14] and the rough surface [$b(1)$= 0.58, $s(1)$= 0.14].   
} 

\subsection{Plasmons  in the  core level spectra of a fully oxidized Al surface}
In  Fig.~S3(a) of SM~\cite{supple}, a fully oxidized  Al crystal surface, shows   oxide and  metal-related components in the Al 1$s$ spectrum.    In spite of being fully oxidized, the metal peak is dominant in normal emission compared to the oxide-related peak at \srb{2.7} eV higher $E$  
~because of the large electron mean free path in HAXPES (80\,\AA\, for Al 1$s$ at 6 keV~\cite{Tanuma}).  The thickness of the oxide film  estimated  from the relative intensity variation of the  metal- and the oxide-related peaks  with $\theta$~\cite{Stroheimer90} turn out to be  48$\pm$2\,\AA.  In   normal emission, the  metal-related bulk plasmon is observed at 15.3 eV loss energy with  $b(1)$= 0.5.  At the grazing emission in  Fig.~S3(b) of SM, although the oxide-related peak is highly enhanced in intensity (shown truncated),  surprisingly, the 1\op\, corresponding to the metal underneath  is still observed with the same \op\ of 15.3 eV and  a   $b(1)$ value of  0.6.  Both these values of $b(1)$ in oxidized Al are close to that of the clean Al metal [$b(1)$= 0.6, Fig.~\ref{bulkplasmon}(b)].    Thus, the normalized intensity of the bulk plasmon is  hardly   affected by the oxide layer formed on the  surface.  This is because both the metal's main peak and the bulk plasmon are similarly attenuated by the oxide layer. However, in contrast, the surface plasmon is completely attenuated  because the metal surface is fully covered by the  oxide film.  

 \subsection{Multiple bulk plasmons in the valence band spectra of Al and Mg}

\begin{figure*}[tb]
	\includegraphics[width=1.2\textwidth,keepaspectratio,trim={1cm 0.5cm 1.8cm 0.1cm},clip]{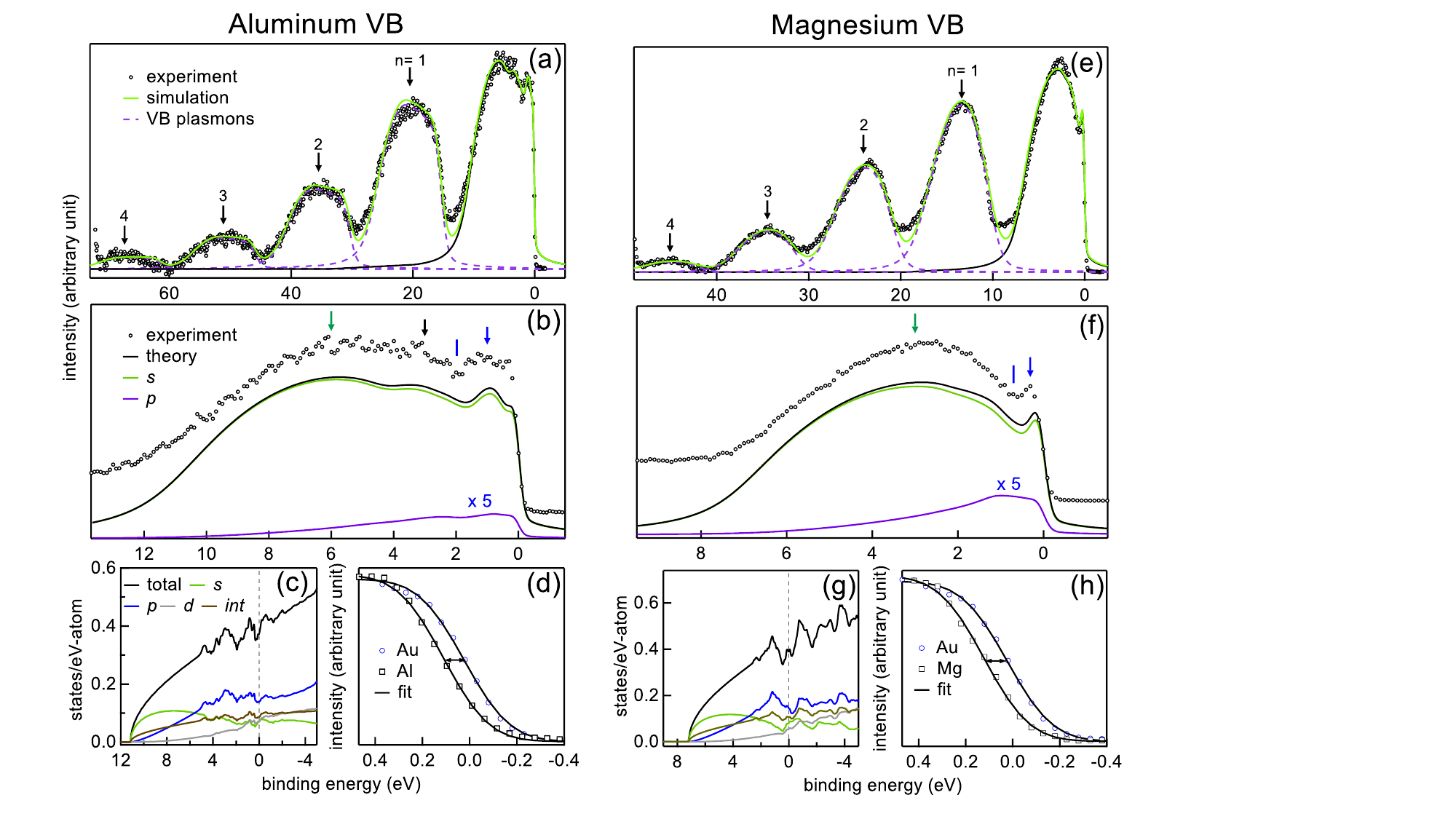}
	\caption{ HAXPES valence band (VB) of (a) Al and (e) Mg metals along with the bulk plasmon loss peaks ($n$\op\  up to $n$= 4  marked by arrows) measured in normal emission after an  inelastic background subtraction. The simulated plasmon line shapes are shown by violet dashed curves, and their addition is shown by a black curve.   (b) Al and (f) Mg  VBs are compared with the calculated VB (black curve), 
		~the $s$ and $p$ partial contributions are shown.   The  calculated VBs are shown overlaid in panels \textbf{a, e} and are  used to simulate the plasmons (see text).  The  DFT calculated (c) Al and (g) Mg total DOS, $s$, $p$ and $d$  PDOS, and the interstitial ($int$) contribution.  The Fermi edges of (d) Al and (h) Mg have been compared to  Au  (black lines are fit to the experimental data); the horizontal double-sided arrow shows the recoil shift.}
	\label{vbplasmon}
\end{figure*}

The HAXPES valence band (VB) spectrum of Al metal, along with the loss region shows, multiple bulk plasmons  $n$\op\ up to $n$= 4 (black arrows) [Fig.~\ref{vbplasmon}(a)]. \srb{The data were recorded for a scraped surface because, for the Ar$^{+}$ ions sputtered surface, the Ar 3$p$ signal arising from implanted Ar bubbles~\cite{Biswas04,Dhaka08} appears at 9.3 eV and interferes with the VB, as shown in Fig.~S4 of SM.}  In Fig.~\ref{vbplasmon}(b), the   theoretical  VB -- calculated using the  PDOS shown in Fig.~\ref{vbplasmon}(c) -- exhibits excellent agreement with the experiment:  All the features, such as a peak near the Fermi level at $E$= 1 eV (blue arrow), a dip at 2 eV (blue tick), a weak feature centered at 3 eV  (black arrow), and the broad hump around 6 eV  (green arrow), are visible at nearly identical energies and possess similar relative intensities.  The VB is largely  dominated by the $s$-like states because  of their large photoemission cross-section  for hard x-rays~\cite{Kobayashi09,Fadley2010,Sadhukhan23,Bhattacharya23}. The small  contribution from $p$-like states  is rather featureless.  
~It should be noted  that the interstitial ($int$) contribution to the total DOS  is significant because a substantial part of the electron density lies outside the atomic sphere that is used to calculate the PDOS [Fig.~\ref{vbplasmon}(c)]. Another point to note is that the PDOS  of Al previously calculated using the KKR method~\cite{Leonard78} differs considerably from our calculation [e.g., compare Fig.~\ref{vbplasmon}(c) with the inset of Fig.~S5 in SM~\cite{supple}].  We used the PDOS from Ref.~\onlinecite{Leonard78} to calculate the VB and find that it  differs significantly from the experiment (Fig.~S5 of SM~\cite{supple}): a peak at $\sim$5 eV is  evident. This is absent in the experimental spectrum, while the dip at $\sim$2 eV (blue tick) is overestimated. These observations show   the shortcomings of the previous calculation, which  is possibly related to unphysical fluctuations on the lower energy side and was performed with constant potential outside the  Muffin tin sphere~\cite{Leonard78}.

In the case of Mg  metal, the VB spectrum also shows multiple bulk plasmons  $n$\op\ up to $n$= 4  with  \op$\sim$10.5~eV [black arrows in Fig.~\ref{vbplasmon}(e)].   The VB, shown on an expanded scale in Fig.~\ref{vbplasmon}(f), exhibits excellent agreement with the theoretical  VB, which is calculated using the  PDOS shown in Fig.~\ref{vbplasmon}(g). An earlier augmented plane wave calculation~\cite{Freeman76} of  the PDOS for Mg agrees well with  our calculation.   
~A peak close to the Fermi level at 0.3 eV (blue arrow), a dip at 0.7 eV (blue tick), and a broad hump around 3 eV (green arrow) are visible at similar energies and  exhibit comparable relative intensities [Fig.~\ref{vbplasmon}(f)].  Here also, as in the case of Al, the VB is largely  dominated by the $s$-like states. 

\begin{figure}[tb]
	\includegraphics[width=0.7\textwidth,keepaspectratio,trim={1cm 0cm 1.5cm 0cm},clip]{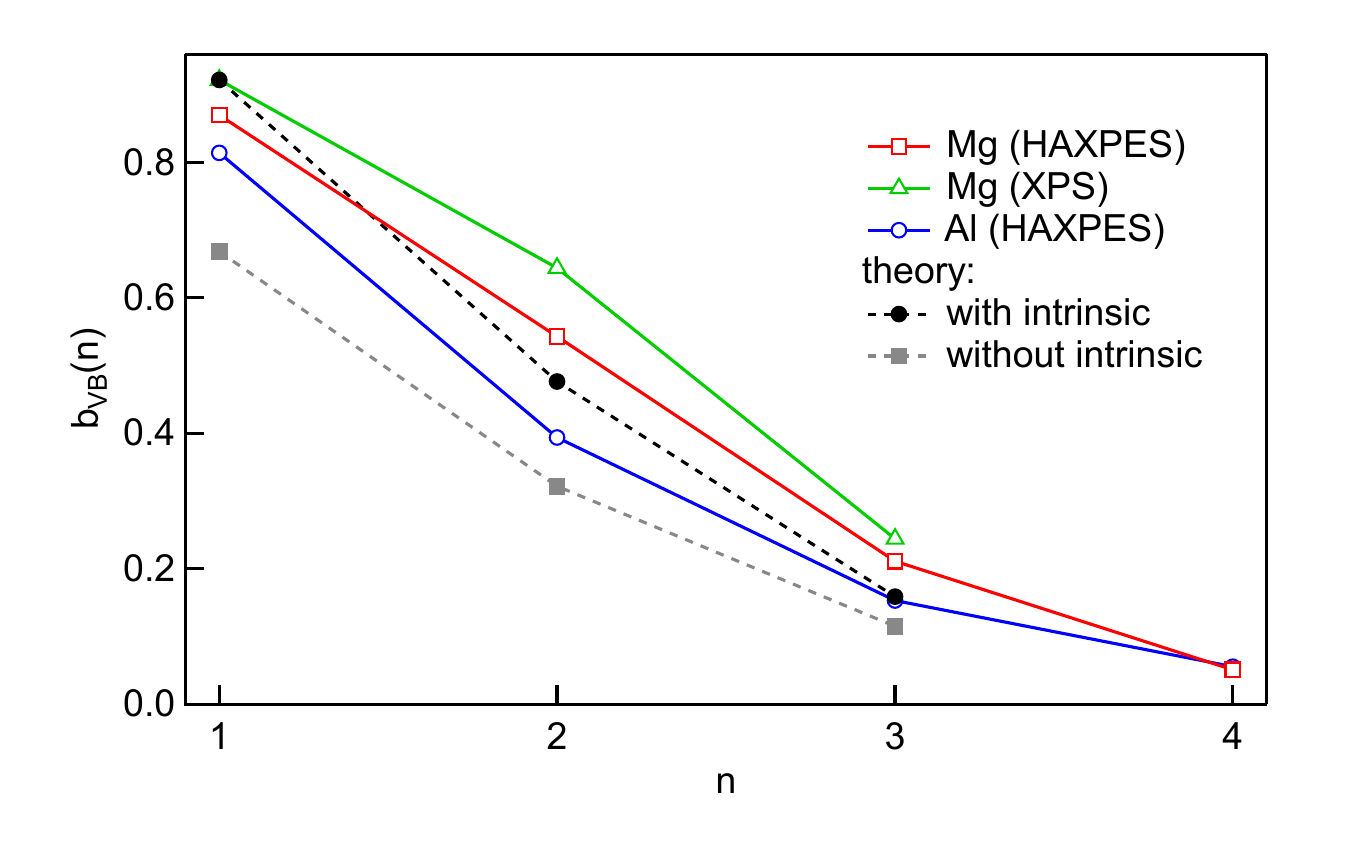}
	\caption{$b_{\rm VB}(n)$ as a function of $n$ for Al and Mg HAXPES VB compared with those extracted from Mg XPS VB~\cite{Hoechst78} and the theoretical calculation by Penn~\cite{Penn78} for Mg with intrinsic and without intrinsic (i.e., only extrinsic) plasmon.}
\label{vbn}
\end{figure}

 The plasmon loss spectra of VB have been calculated  for Mg and Na by Penn~\cite{Penn78}, which drew on the work by Langreth~\cite{Langreth70}. Unlike  the core level  intrinsic plasmon that is created by a localized core hole in the final state,  the photo-hole in the VB is delocalized. So, the mechanism of  intrinsic plasmon production in the VB was proposed to be  related to electron-electron interaction:  a second electron creates the intrinsic plasmon  via its interaction with the  conduction electrons and scatters into the hole left by the photoexcited electron~\cite{Penn78}.   Although  energy dependence and interference effect were not considered in theory~\cite{Penn78}, remarkable agreement was obtained with the XPS spectra~\cite{Hoechst78} only when  both extrinsic and intrinsic plasmons were considered. On the other hand,  the plasmon intensities were significantly underestimated  if the intrinsic plasmon was not considered. This established the presence of  intrinsic plasmons  in the XPS VB plasmon  in  nearly free-electron metals~\cite{Penn78}.

To compare theory and experimental results,   the  shapes of the $n^{\rm th}$ VB plasmons for both Al and Mg have been simulated 
~by shifting the calculated VB by $n$\op\ towards higher $E$, and allowing the  intensity and an additional Gaussian broadening to vary [Figs.~\ref{vbplasmon}(a,e)]. We find that unlike the core level plasmons, asymmetric broadening is not required, and the FWHM of the Gaussian remains nearly unchanged with $n$ (2 eV and 2.5  eV for Al and Mg, respectively). 
 The $b_{\rm VB}(n)$ values of Mg  have also been extracted for  the theoretical calculations by Penn~\cite{Penn78} for XPS and the the XPS spectrum published in the literature~\cite{Hoechst78}   (Fig.~S6 of SM~\cite{supple}). This has been done by shifting the calculated XPS VB by $n$\op\ and broadening it by a Gaussian,  as  discussed above.  Figure~S6 of SM~\cite{supple} shows good agreement between the calculated and experimental XPS VB. The former is  obtained from the PDOS in Figs.~\ref{vbplasmon}(c,g) using the photoemission cross-sections for AlK$_{\alpha}$ (1486.6 eV) radiation~\cite{Trzhaskovskaya18}. 
 
 Figure~\ref{vbn} shows  $b_{\rm VB}(n)$, which is the area of the $n^{\rm th}$ plasmon  divided by the area of the VB. \srb{$b_{\rm VB}(n)$ for  Mg and Al HAXPES VB are close, as shown by the red and blue curves.}
~$b_{\rm VB}(1)$  for Mg  HAXPES  (0.87) and XPS (0.9) are quite similar to each other and  to the theoretical  value (0.9) of Mg  that includes the  intrinsic and extrinsic plasmons (black filled circle-dashed line).  On the other hand, if intrinsic plasmon is not considered, the contribution of only the extrinsic process is significantly less (0.6) \srb{(gray filled square-dashed line).  Similarly, $b_{\rm VB}(2)$ and $b_{\rm VB}(3)$ for both Mg HAXPES and XPS  are closer to the theoretical values that include both  intrinsic and extrinsic plasmons.}   Thus, our data   provide  evidence of  the intrinsic plasmon production in the VB related plasmons of Mg  measured by HAXPES. 

 \subsection{Recoil effect in  the valence band spectra of  Mg}

The recoil effect is well known in the HAXPES core level spectra of light elements. When a photoelectron of mass $m$ is emitted with a large kinetic energy ($E_{kin}$) \srb{from an atom of mass $M$}, the photoelectron delivers a recoil energy ($E_R$) to that atom, which is given by \srb{$E_R$$\approx$$\frac{m}{M}$$\times E_{kin}$~\cite{Takata08,Fadley2010}}. The photoelectrons lose this energy, $E_R$, resulting in a recoil shift towards higher $E$. This effect has also been observed in the VB  of light metals such as Al (atomic mass 27 u) 
~through a shift of  the Al metal Fermi edge compared to that of Au~\cite{Takata08}. Au,  being of high mass (197 u),  hardly exhibits this effect.  Recently, such a recoil shift in the VB has been reported in a light complex metallic alloy $\beta$-Al$_3$Mg$_2$~\cite{Singh22}. 
~Here in Fig.~\ref{vbplasmon}(d),  the recoil shift of Al is, however, smaller (85 meV) compared to 120 meV reported  using 7.94 keV~\cite{Takata08}. This can be explained by the proportionality of the recoil shift with $E_{kin}$ i.e., $\frac{6~\rm keV}{7.94~\rm keV} \times$120 meV= 90.7 meV, which is close to 85 meV. 
~Due to marginally smaller mass  of Mg (24.3 u)  compared to Al, the recoil shift is  found to be slightly larger [90 meV,  Fig.~\ref{vbplasmon}(h)].\\

\section{Conclusion}
Multiple bulk plasmons ($n$\op) in \srb{the valence band (up to $n$= 4) and core level spectra (up to $n$= 13)}  of free electron metals such as Al and Mg  have been studied by hard x-ray photoelectron spectroscopy (HAXPES).  
~Based on earlier theoretical works~\cite{Langreth74,Langreth71,Chang72, Chang72a, Penn78}, estimates of the extrinsic, intrinsic, and interference processes that contribute to the core level plasmon intensities  are obtained from exhaustive multi parameter curve fitting. The probabilities of the extrinsic and intrinsic  processes in the core level plasmons are nearly similar for the two free electron metals,  $\alpha$  and $\beta$ being  0.54$\pm$0.05 and 0.18$\pm$0.1, respectively. 
 These values imply that, as $n$ increases, the intrinsic plasmon contribution decreases  while the interference plasmon changes  from negative to zero.  Despite the larger inelastic mean free path in HAXPES, the intrinsic, extrinsic, and interference processes  remain nearly unchanged across a broad kinetic energy range  (1–6 keV). This finding is consistent with previous theoretical research~\cite{Shinotsuka08}.  The 1\op\, line shape is  more asymmetric because of the presence of  sizable  intrinsic and interference terms.   The 1\os\, surface plasmon has been identified in  both Al and Mg. Its intensity increases  significantly in the grazing emission compared to the normal emission.  In the case of   oxidized aluminum, the surface plasmon is completely suppressed, while the bulk plasmon intensity is almost unaffected compared to the metal's main peak.  

Multiple plasmons are observed in the loss region of the Al and Mg valence bands; the shape of the VB is in excellent agreement with the DFT-based calculated valence band.  \srb{Comparison with the theoretical calculation~\cite{Penn78} after extracting the relative VB plasmon intensities establish the existence of the intrinsic  plasmon, besides the extrinsic plasmon.} The recoil shift observed  in both Al and Mg VB is related to the kinetic energy of the photoelectron and the mass of the atom.

\section{Acknowledgment}
The HAXPES experiments were carried out at PETRA III of Deutsches Elektronen-Synchrotron, a member of Helmholtz-Gemeinschaft Deutscher Forschungszentren. Financial support by the Department of Science and Technology, Government of India, within the framework of the India@DESY collaboration is gratefully acknowledged. We are thankful to  C. Schlueter, R. Choubisa, and D. S\'ebilleau  for support and encouragement. The Computer division of the Raja Ramanna Centre for Advanced Technology is thanked for installing the DFT codes and providing support throughout. We thank  I. Schostak for  skillful technical support.\\

\noindent $^*$e-mail address: barmansr@gmail.com


\begin{thebibliography}{82}
\bibitem [{Lor()}]{Lorentzian}%
	\BibitemOpen
	\href@noop {} {\bibinfo  {journal} {$\frac {I_o} {1+[(E_K-E_K^p) /
				\Gamma(E)]^2}$, where the half width at half maximum, $\Gamma(E)$= $\Gamma_R$
			when $E_K > E_K^p$ ($E_K^p$ is the peak position in the kinetic energy scale,
			$E_K$) and $\Gamma(E)$= $\Gamma_L$ when $E_K < E_K^p$, and $I_o$ is the
			intensity}\ }\BibitemShut {NoStop}%
\bibitem [{bet()}]{beta}%
	\BibitemOpen
	\href@noop {} {\bibinfo  {journal} {$P_{\rm int}(n)$, the probability of
			intrinsic process for the $n^{\rm th}$ plasmon is given by $
			e^{-\beta}\frac{\beta^n}{n!}$}\ }\BibitemShut {NoStop}%
\end{thebibliography}

%

\end{document}